\documentclass{article}

\usepackage{graphicx} 
\usepackage{subfigure}

\usepackage{natbib}

\usepackage{algorithm}
\usepackage{algorithmic}

\usepackage{times}

\usepackage{hyperref}  

\usepackage[accepted]{icml2012}

\usepackage{xspace}
\usepackage{amsmath}
\usepackage{amssymb}
\usepackage{xcolor}

\icmltitlerunning{Latent Multi-group Membership Graph Model}

\begin{document}

\twocolumn[
\icmltitle{Latent Multi-group Membership Graph Model
           }

\icmlauthor{Myunghwan Kim}{mykim@stanford.edu}
\icmladdress{Stanford University,
            Stanford, CA 94305, USA}
\icmlauthor{Jure Leskovec}{jure@cs.stanford.edu}
\icmladdress{Stanford University,
            Stanford, CA 94305, USA}

\icmlkeywords{network model, latent group, node features, multi-group}

\vskip 0.3in
]

\newcommand{\fix}{\marginpar{FIX}}
\newcommand{\new}{\marginpar{NEW}}
\newcommand{\xhdr}[1]{\vspace{1.7mm}\noindent{{\bf #1.}}}

\newcommand{\denselist}{\itemsep -4pt\topsep -20pt\partopsep -12pt}
\newcommand{\rev}[1]{{#1}}
\newcommand{\eg}{\textit{e.g.}\xspace}
\newcommand{\ie}{\textit{i.e.}\xspace}
\newcommand{\model}{\emph{LMMG}\xspace}
\newcommand{\fullmodel}{\emph{Latent Multi-group Membership Graph}\xspace}
\newcommand{\Topic}{Group\xspace}
\newcommand{\topic}{group\xspace}
\newcommand{\magfit}{MAG-FIT}
\newcommand{\hide}[1]{}
\newcommand{\LONGER}{\rev{Appendix}}

\begin{abstract}
We develop the \fullmodel (\model) model, a model of networks with rich node feature structure. In the \model model, each node belongs to multiple groups and each latent group models the occurrence of links as well as the node feature structure. The \model can be used to summarize the network structure, to predict links between the nodes, and to predict missing features of a node. We derive efficient inference and learning algorithms and evaluate the predictive performance of the \model on several social and document network datasets.

\end{abstract}

\section{Introduction}
\label{sec:intro}

Network data, such as social networks of friends, citation networks of documents, and hyper-linked networks of webpages, play an increasingly important role in modern machine learning applications. Analyzing network data provides useful predictive models for recommending new friends in social networks~\cite{backstrom11supervisedrw} or scientific papers in document networks~\rev{\cite{nallapati08lda,chang09rtm}}.

Research on networks has focused on various models of network link structure. Latent variable models~\cite{airoldi07blockmodel,hoff02latent,kemp06irm} decompose a network according to hidden patterns of connections between the nodes, while models based on Kronecker products~\cite{jure10kronecker,mh12mag,mh11kronem} accurately model the global network structure. Though powerful, these models account only for the structure of the network, while ignoring observed features of the nodes. For example, in social networks users have profile information, and in document networks each node also contains the text of the document that it represents. Such models can find patterns which account for the connections between nodes, but they cannot account for the node features.

\hide{
}

Node features along with the links between them provide rich and complementary sources of information and should be used simultaneously for uncovering, understanding and exploiting the latent structure in the data. In this respect, we develop
\rev{a new network model}
\rev{considering}
both the emergence of links of the network and the structure of node features such as user profile information or text of a document.

Considering both sources of data, links and node features, leads to more powerful models than those that only consider links. For example, given a new node with a few of its links, traditional network models provide a predictive distribution of nodes to which it might be connected.
However,
\rev{to predict links of a node,}
our model does not need to see any links of a node. It can predict links using only node's features. For example, we can suggest user's friendships based only on the profile information, or recommend hyperlinks of a webpage based only on its textual information. Moreover, given a new node and its links, our model also provides a predictive distribution of node features. This can be used to predict features of a node given its links or even predict missing or hidden features of a node given its links. For example, in our model user's interests or keywords of a webpage can be predicted using only the connections of the network. Such predictions are out of reach for traditional models of networks.

\begin{figure}[t]
\centering
\includegraphics[width=0.4\textwidth]{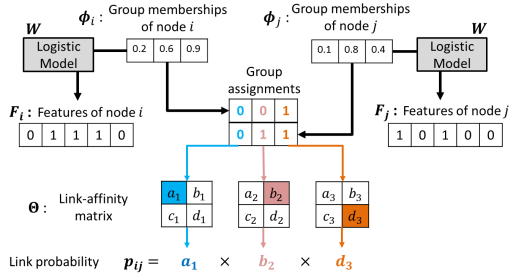}
\vspace{-2mm}
\caption{\fullmodel model. A node belongs to multiple latent groups at once. Based on group memberships features of a node are generated using a logistic model. Links are modeled via link-affinity matrices which allows for rich interactions between members and non-members of groups.}
\label{fig:example}
\vspace{-4mm}
\end{figure}

We develop a \fullmodel (\model) model of networks that explicitly ties nodes into groups of shared features and linking structure (Figure~\ref{fig:example}). Nodes belong to multiple latent groups and the occurrence of each node feature is determined by a logistic model based on the \topic~memberships of the given node. Links of the network are then generated via link-affinity matrices. Each link-affinity matrix $\Theta_i$ represents a table of link probabilities, and
an appropriate entry of $\Theta_i$ is chosen
based on whether or not a pair of nodes share the membership in group $i$.
We derive effective algorithms for model parameter estimation and prediction. We study the performance of \model on real-world social and document networks. We investigate the predictive performance on three different tasks: link prediction, node feature prediction, and supervised node classification. The \model provides significantly better performance on all three tasks than natural alternatives and the current state of the art.

\section{\model Model Formulation}
\label{sec:model}
The \fullmodel (\model) model is a model of a (directed or undirected) network and nodes which have categorical features. Our model contains two important ingredients or innovations (See Figure~\ref{fig:example}).

First, the model assigns nodes to latent groups and allows nodes to belong to multiple groups at once. In contrast to multinomial models of group membership \cite{airoldi07blockmodel,chang09rtm}, where the membership of a node is shared among the groups (the probability over group memberships of a node sums to 1), we model group memberships as \rev{a series of} Bernoulli random variables ($\phi_{i}$ in Figure~\ref{fig:example}), which indicates that nodes in our model can truly belong to multiple groups. Hence, in contrast to multinomial topic models,
a higher probability of  node membership to a group does not necessarily to lower probability of membership to some other group in the \model.

Second, for modeling the links of the network,
each group $k$
has associated
a link-affinity matrix ($\Theta$ in Figure~\ref{fig:example}). Each link-affinity matrix represents a table of link probabilities given that a pair of nodes belongs or does not belong to group $k$. Thus, depending on the combination of the memberships of nodes to group \rev{$k$}, an appropriate element of \rev{$\Theta_k$} is chosen. For example, the entry $(0,0)$ of $\Theta_k$ captures the link-affinity when none of the nodes belongs to group \rev{$k$}, while $(1,0)$ stores the link-affinity when first node belongs to the group but the second does not. As we will later show \rev{that} this allows for rich flexibility in modeling the links of the network as well as for uncovering and understanding the latent structure in the network data.

\begin{figure}[t]
\centering
\includegraphics[width=0.4\textwidth]{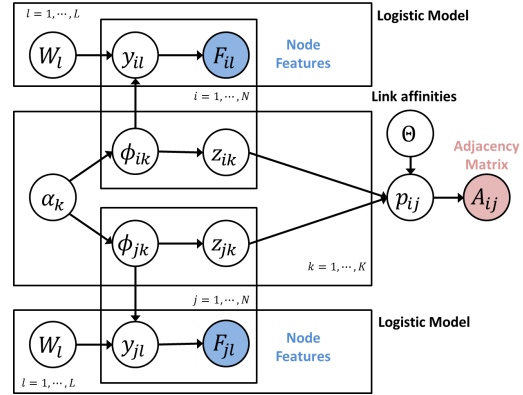}
\vspace{-2mm}
\caption{Plate model representation of \model~model.}
\label{fig:platemodel}
\vspace{-2mm}
\end{figure}

Now we formalize the \model model illustrated in Figure~\ref{fig:platemodel} and describe it in a generative way. Formally, each node $i = 1, 2, \cdots, N$ has a real-valued \topic membership $\phi_{ik} \in [0, 1]$ for each \topic $k = 1, 2, \cdots, K$. $\phi_{ik}$ represents the probability that node $i$ belongs to \topic $k$. Assuming the Beta distribution parameterized by $\alpha_{k1}, \alpha_{k2}$ as a prior distribution of \topic membership $\phi_{ik}$,
we model the latent \topic assignment $z_{ik}$ for each node as follows:
\begin{align}
\label{eq:modeltopic}
\phi_{ik} & \sim \mathrm{Beta}(\alpha_{k1}, \alpha_{k2}) ~~~\nonumber \\
z_{ik} & \sim \mathrm{Bernoulli}(\phi_{ik}) ~~~\mathrm{for}~k = 1, 2, \cdots, K \, .
\end{align}
Since each group membership $z_{ik}$ of a node is independent, a node can belong to multiple groups simultaneously.

The \topic~memberships of a node affect both node features and its links. With respect to node features, we limit our focus to binary-valued
\rev{features}
and use a logistic function to model the occurrence of node's features based on the groups it belongs to.
For each feature $F_{il}$ of node $i$ ( $l = 1, \cdots, L$ ), we consider a separate logistic model where we regard \topic memberships $\phi_{i1}, \cdots, \phi_{iK}$ as input features of the model. In this way, the logistic model represents the relevance of each \topic membership to the presence of a node feature.
For convenience, we refer to the input vector of node $i$ for the logistic model as $\phi_{i} = [\phi_{i1}, \cdots, \phi_{iK}, 1]$, where $\phi_{i(K+1)} = 1$ \rev{represents} the intercept term. \rev{Then,}
\begin{align}
\label{eq:modelattr}
y_{il} & = \frac{1}{1 + \exp(- w_{l}^{T} \phi_{i})} \nonumber \\
F_{il} & \sim \mathrm{Bernoulli}(y_{il}) ~~~\mathrm{for}~l = 1, 2, \cdots, L
\end{align}
where $w_{l} \in \mathbb{R}^{K+1}$ is the logistic model parameter for the $l$-th node feature.
The value of each $w_{lk}$ indicates the contribution of \topic $k$ to the presence of node feature $l$.

\begin{figure}
\centering
\begin{tabular}{ccc}
\includegraphics[width=0.10\textwidth]{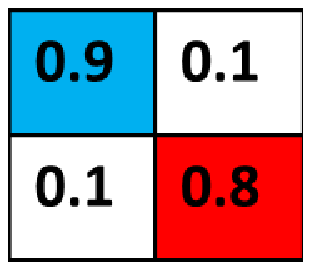} &
\includegraphics[width=0.10\textwidth]{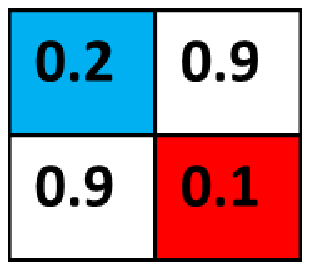} &
\includegraphics[width=0.10\textwidth]{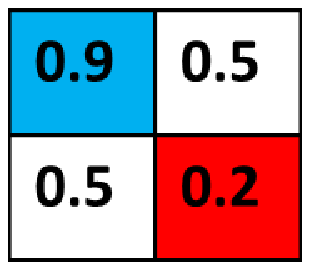} \\
(a) Homophily & (b) Heterophily & (c) Core-periphery
\end{tabular}
\caption{Link structures modeled by link-affinity matrices.}
\vspace{-4mm}
\label{fig:netstructure}
\end{figure}

In order to model the links of the network, we build on the idea of the Multiplicative Attributes Random Graph (MAG) model~\cite{mh12mag}. Here each latent \topic $k$ has associated a link-affinity matrix $\Theta_{k} \in [0, 1]^{2 \times 2}$. Each entry of the link-affinity matrix indicates a tendency of linking between a pair of nodes depending on whether they belong to the \topic $k$ or not. In other words, given the \topic assignments $z_{ik}$ and $z_{jk}$ of nodes $i$ and $j$, $z_{ik}$ ``selects'' a row and $z_{jk}$ ``selects'' a column of $\Theta_{k}$ and so that the linking tendency from node $i$ to node $j$ is captured by $\Theta_{k}[z_{ik}, z_{jk}]$.
After acquiring such link-affinities from all the groups, we define the link probability $p_{ij}$ as the product of the link-affinities. Therefore, based on latent \topic assignments and link-affinity matrices,
we determine each entry of the adjacency matrix $A \in \{0, 1\}^{N \times N}$ of the network as follows:
\begin{align}
\label{eq:modelnet}
p_{ij} & =  \prod_{k} \Theta_{k} [z_{ik}, z_{jk}] \nonumber \\
A_{ij} & \sim \mathrm{Bernoulli}( p_{ij} ) ~~~\mathrm{for}~i, j = 1, 2, \cdots N \, .
\end{align}
%
\rev{The network model parameter $\Theta_{k}$ represents} the link affinity with respect to the particular \topic $k$. The model offers flexibility in a sense that we can represent many types of linking structures. \rev{In} Figure~\ref{fig:netstructure},
by varying the link-affinity matrix, the model can capture heterophily (love of the different), homophily (love of the same), or core-periphery structure.
This way the affinity matrix allows us to discover the effects of node features on links of the network.

\rev{The node feature and the network models are connected via \topic~memberships $\phi_{i}$.}
For instance, suppose that $w_{lk}$ is large for some feature $l$ and topic $k$.
Then, as the node $i$ belongs to topic $k$ with high probability ($\phi_{ik}$ is close to $1$),
the feature $l$ of node $i$, $F_{il}$, is more likely to be $1$. By modeling group memberships using \rev{multiple} Bernoulli random variables (instead of using multinomial distribution~\cite{airoldi07blockmodel,chang09rtm}), we achieve greater modeling flexibility which allows for 
making predictions
about links given features and features given links. In Section \ref{sec:experiments}, we empirically demonstrate that the \model outperforms
traditional models on these tasks.

Moreover, if we divide the nodes of the network into two sets depending on the membership to \topic $k$, then 
\rev{we can discover how members of group $k$ link to other members as well as non-members of $k$, based on the structure of $\Theta_{k}$.} 
For example, \rev{when} $\Theta_{k}$ has large values on diagonal entries like in Figure~\ref{fig:netstructure}(a),
\rev{members or non-members}
are likely to link among themselves, while there is low affinity for links between members and non-members. Figure~\ref{fig:netstructure}(b) captures exactly the opposite behavior where links are most likely between members and non-members. While the core-periphery structure is captured by link-affinity matrix in Figure~\ref{fig:netstructure}(c) where nodes that share group memberships (the ``core'') are most likely to link, while nodes in the periphery are least likely to link among themselves.


\section{Inference, Estimation and Prediction}
\label{sec:algorithm}
\newcommand{\LPH}{\mathcal{L}_{\phi}}
\newcommand{\LF}{\mathcal{L}_{F}}
\newcommand{\LA}{\mathcal{L}_{A}}
\newcommand{\LZ}{\mathcal{L}}
\newcommand{\PLPH}[1]{\frac{\partial \mathcal{L}_{\phi}}{\partial {#1}}}
\newcommand{\PLF}[1]{\frac{\partial \mathcal{L}_{F}}{\partial {#1}}}
\newcommand{\PLA}[1]{\frac{\partial \mathcal{L}_{A}}{\partial {#1}}}
\newcommand{\PLZ}[1]{\frac{\partial \mathcal{L}}{\partial {#1}}}
\newcommand{\EX}{\mathbb{E}}
\newcommand{\EZM}{\EX_{Z|M}}
\newcommand{\xset}{\char`\\}
\newcommand{\OBS}{\mathbb{O}}
\newcommand{\EZPH}{\EX_{Z \sim \phi}}

We now turn our attention to \model model estimation. Given a set of binary node features $F$ and the network $A$, we aim to find node group memberships $\phi$, parameters $W$ of node feature model, and link-affinity matrices $\Theta$.

\subsection{Problem formulation}
\rev{When} the node features $F = \{F_{il}: i = 1, \cdots, N, ~~l = 1, \cdots, L\}$ and the adjacency matrix $A \in \{0, 1\}^{N \times N}$ are given, 
we aim to find the \topic memberships $\phi = \{\phi_{ik}: i = 1, \cdots, N, ~~k = 1, \cdots, K\}$, the logistic model parameters $W =  \{w_{lk}: l = 1, \cdots, L, ~~k = 1, \cdots, K + 1\}$, and the link-affinity matrices $\Theta = \{\Theta_{k}: k = 1, \cdots K\}$. We apply the maximum likelihood estimation, which finds the optimal values of $\phi$, $W$, and $\Theta$ so that they maximize the likelihood $P(F, A, \phi | W, \Theta, \alpha)$ where $\alpha$ represents hyper parameters, $\alpha = \{(\alpha_{k1}, \alpha_{k2}): k = 1, \cdots, K\}$,
\rev{for the Beta prior distributioins}. In the end, we aim to solve
\begin{align}
\label{eq:problem1}
\max_{\phi, W, \Theta} \log P(F, A, \phi | W, \Theta, \alpha) \, .
\end{align}
Now we 
\rev{compute}
the objective function in the above optimization problem. Since the \model independently generates $F$ and $A$ given \topic memberships $\phi$, we decompose the log-likelihood $\log P(F, A, \phi | W, \Theta, \alpha)$ as follows:
\begin{align}
\label{eq:lik}
& \log P(F, A, \phi | W, \Theta, \alpha)  \nonumber \\
& \, = \log P(F | \phi, W) + \log P(A | \phi, \Theta) + \log P(\phi | \alpha) \, .
\end{align}
Hence, to compute $\log P(F, A, \phi| W, \Theta, \alpha)$, we separately calculate each term of Equation~(\ref{eq:lik}). We obtain $\log P(\phi | \alpha)$ and $\log P(F | \phi, W)$ from Equations~(\ref{eq:modeltopic}) and (\ref{eq:modelattr}):
\begin{align}
& \log P(\phi | \alpha) = \sum_{i, k} (\alpha_{k1} - 1) \log \phi_{ik} \nonumber \\
& \quad \quad \quad \quad \quad \quad \quad \quad \quad + \sum_{i, k} (\alpha_{k2} - 1) \log (1-\phi_{ik}) \nonumber \\
& \log P(F | \phi, W) = \sum_{i, l} F_{il} \log y_{il} + (1 - F_{il}) \log (1 - y_{il}) \nonumber
\end{align}
where $y_{il}$ is defined in Equation~(\ref{eq:modelattr}).

With regard to the second term in Equation~(\ref{eq:lik}),
\begin{align}
\label{eq:exact}
\log P(A | \phi, \Theta)
= \log \sum_{Z} P(A|Z, \phi, \Theta) P(Z | \phi, \Theta)
\end{align}
for $Z = \{z_{ik}: i = 1, \cdots, N, ~~k = 1, \cdots, K\}$.
\rev{We note} that $A$ is independent of $\phi$ given $Z$. 
\rev{To exactly calculate $\log P(A | \phi, \Theta)$,}
we thus 
\rev{sum} $P(A|Z, \Theta)P(Z|\phi)$ 
\rev{over every instance of $Z$ given $\Theta$ and $\phi$,} 
but this requires the sum over $2^{NK}$ instances.
\rev{As this exact computation is infeasible,}
we approximate
\rev{$\log P(A | \phi, \Theta)$ using its lower bound obtained }
by applying Jensen's Inequality
\rev{to Equation~(\ref{eq:exact}):}
%
\begin{align}
\label{eq:lbnet}
\log P(A | \phi, \Theta)
& = \log \EX_{Z \sim \phi}\left[ P(A | Z, \Theta) \right] \nonumber \\
& \geq \EX_{Z \sim \phi} \left[ \log P(A | Z, \Theta) \right]
\end{align}
%
Now that we are summing up over $N^2$ terms, the computation of the lower bound is feasible. We thus maximize the lower bound $\LZ$ of the log-likelihood $\log P(A, F, \phi | W, \Theta, \alpha)$. To sum up, we aim to maximize
%
%
\begin{align}
\label{eq:problem2}
\min_{\phi, W, \Theta} - (\LPH + \LF + \LA) + \lambda |W|_{1} 
\end{align}
where $\LPH = \log P(\phi | \alpha), \LF = \log P(F | \phi, W)$, and $\LA = \EX_{Z \sim \phi} \left[ \log P(A | Z, W) \right]$. 
To
avoid overfitting, we regularize the objective function by the L1-norm of $W$.



\subsection{Parameter estimation}

To solve the problem in Equation (\ref{eq:problem2}), we alternately update the \topic memberships $\phi$, the model parameters $W$, and $\Theta$. Once $\phi$, $W$, and $\Theta$ are initialized, we first update the \topic memberships $\phi$ to maximize $\LZ$ with fixing $W$ and $\Theta$. We then update the model parameters $W$ and $\Theta$ to 
minimize the function $(-\LZ + \lambda |W|_{1})$ in Equation (\ref{eq:problem2}) 
by fixing $\phi$.
Note that $\LZ$ is decomposed into $\LA$, $\LF$, and $\LPH$. 
Therefore, when updating $W$ and $\Theta$ given $\phi$, we separately maximize the corresponding log-likelihoods $\LF$ and $\LA$.
We repeat this alternate updating procedure until the solution converges.
In the following we describe the details.

\hide{
\subsubsection{Initialization}
Since the objective function in Equation~(\ref{eq:problem2}) is non-convex, the final solution might be dependent on the initial values of $\phi$, $W$, and $\Theta$.
For reasonable initialization, as the node features $F$ are given, we run the Singular Vector Decomposition (SVD) by regarding $F$ as an $N \times L$ matrix
and obtain the singular vectors corresponding to the top $K$ singular values.
By taking the top $K$ components, we can approximate the node features $F$ over $K$ latent dimensions.
We thus assign the $l$-th entry of the $k$-th right singular vectors multiplied by the $k$-th singular value into $w_{lk}$ for $l = 1, \cdots, L$ and $k = 1, \cdots, K$.
We also initialize each \topic membership $\phi_{ik}$ based on the $i$-th entry of the $k$-th left singular vectors.
This approximation can in particular provide good enough initial values
when the top $K$ singular values dominate the others.
In order to obtain the sparse model parameter $W$,
we reassign $0$ to $w_{lk}$ of small absolute value such that $|w_{lk}| < \lambda$.

Finally, to initialize the link-affinity matrices $\Theta$, we introduce the following way.
When initializing the $k$-th link-affinity matrix $\Theta_{k}$,
we assume that the \topic other than \topic $k$ has nothing to do with network structure, \ie~every entry in the other link-affinity matrices has the equal value.
Then, we compute the ratio between entries $\Theta_{k}[x_{1}, x_{2}]$ for $x_{1}, x_{2} \in \{0, 1\}$ as follows:
\begin{align*}
\Theta_{k}[x_{1}, x_{2}] \propto \sum_{i, j: A_{ij} = 1} \EZPH P[z_{ik} = x_{1}, z_{ik} = x_{2}]
\end{align*}
As the \topic membership $\phi$ is initialized above and $z_{ik}$ and $z_{jk}$ are independent of each other,
we are able to compute the ratio between entries of $\Theta_{k}$.
After computing the ratio between entries for each link-affinity matrix, we adjust the scale of the link-affinity matrices
so that the expected number of edges in the MAG model is equal to the number of edges in the given network, \ie~$\sum_{i, j} p_{ij} = \sum_{i, j} A_{ij}$.
}

\xhdr{Update of \topic memberships $\phi$}
Now we focus on the update of \topic membership $\phi$ given the model parameters $W$ and $\Theta$. We use the coordinate ascent algorithm which updates each membership $\phi_{ik}$ by fixing the others so to maximize the lower bound $\LZ$. By computing the derivatives of $\LPH$, $\LF$, and $\LA$ we apply the gradient method to update each $\phi_{ik}$:
%
\begin{align}
\label{eq:updatephiw}
\PLPH{\phi_{ik}} &= \frac{\alpha_{k1} - 1}{\phi_{ik}} - \frac{\alpha_{k2} - 1}{1-\phi_{ik}} \nonumber \\
\PLF{\phi_{ik}} &= \sum_{l} (F_{il} - y_{il}) w_{lk} \nonumber \\
\PLA{\phi_{ik}} &= \EZPH \left[ \sum_{j: A_{ij} = 1} \frac{\partial \log p_{ij}}{\partial \phi_{ik}} + \sum_{j: A_{ij} = 0} \frac{\partial \log (1-p_{ij})}{\partial \phi_{ik}} \right. \nonumber \\
& + \left. \sum_{j: A_{ji} = 1} \frac{\partial \log p_{ji}}{\partial \phi_{ik}} + \sum_{j: A_{ji} = 0} \frac{\partial \log (1-p_{ji})}{\partial \phi_{ik}} \right]
\end{align}
where $F_{il}$ is either $0$ or $1$, and $y_{il}$ and $p_{ij}$ is respectively defined in Equation~(\ref{eq:modelattr}) and (\ref{eq:modelnet}).
%
Due to the brevity, we describe the details of Equation~(\ref{eq:updatephiw}) in the \LONGER. Hence, by adding up $\PLPH{\phi_{ik}}$, $\PLF{\phi_{ik}}$, and $\PLA{\phi_{ik}}$, we complete computing the derivative of the lower bound of log-likelihood $\PLZ{\phi_{ik}}$ and update the \topic membership $\phi_{ik}$ using the gradient method:
\begin{align}
\label{eq:updatephi}
\phi_{ik}^{new} = \phi_{ik}^{old} + \gamma_{\phi} \left( \PLA{\phi_{ik}} + \PLF{\phi_{ik}} + \PLA{\phi_{ik}} \right)
\end{align}
for a given learning rate $\gamma_{\phi}$.
By updating each $\phi_{ik}$ in turn with fixing the others, we can find the optimal \topic memberships $\phi$ given the model parameters $W$ and $\Theta$.

\xhdr{Update of node feature model parameters $W$}
Now we update the parameters for node feature model, $W$, while \topic memberships $\phi_{ik}$ are fixed. Note that given the \topic membership $\phi$ the node feature model and the network model are independent of each other.
Therefore, finding the parameter $W$ is identical to running the L1-regularized logistic regression given input $\phi$ and output $F$ data
as we penalize the objective function in Equation~(\ref{eq:problem2}) on the L1 value of the model parameter $W$.
We basically use the gradient method to update $W$ but make it sparse by applying the technique similar to \textit{LASSO}:
%
%
\begin{align}
\label{eq:updatew}
\PLF{w_{lk}}  & = \sum_{i} (F_{il} - y_{il}) \phi_{ik} \nonumber \\
w_{lk}^{new} & = w_{lk}^{old} + \gamma_{F} \PLF{w_{lk}} - \lambda(k) \mathrm{Sign}(w_{lk})
\end{align}
if $w_{lk}^{old} \neq 0$ or $|\PLF{w_{lk}}| > \lambda(k)$ where $\lambda(k) = \lambda$ for $k = 1, \cdots, K$ and $\lambda(K+1) = 0$ (\ie, we do not regularize on the intercepts). $\gamma_{F}$ is a constant learning rate.
Furthermore, if $w_{lk}$ crosses $0$  while being updated,
we assign $0$ to $w_{lk}$ as \textit{LASSO} does.
By this procedure, we can update the node feature model parameter $W$ to maximize the lower bound of log-likelihood $\LZ$ as well as to maintain the small number of relevant groups for each node feature.

%
%



\xhdr{Update of network model parameters $\Theta$}
Next we focus on updating network model parameters, $\Theta$, also where the \topic membership $\phi$ is fixed. Again, note that the network model is independent of the node feature model given the \topic membership $\phi$,
so we do not need to consider $\LPH$ or $\LF$. We thus update $\Theta$ to maximize $\LA$ given $\phi$ using the gradient method.
\begin{align}
\nabla_{\Theta_{k}} \LA & \approx \nabla_{\Theta_{k}} \EZPH \left( \sum_{A_{ij}=1} \log p_{ij} + \sum_{A_{ij} = 0} \log (1-p_{ij}) \right) \nonumber \\
\Theta_{k}^{new} & = \Theta_{k}^{old} + \gamma_{A} \nabla_{\Theta_{k}} \LA \nonumber
\end{align}
for a constant learning rate $\gamma_{A}$. We explain the computation of $\nabla_{\Theta_{k}} \EZPH \log p_{ij}$ and $\nabla_{\Theta_{k}} \EZPH \log (1-p_{ij})$ in detail in the \LONGER.

\subsection{Prediction}
With a fitted model, our ultimate goal is to make predictions about new data. In the real-world application, the node features are often missing. Our algorithm is able to nicely handle such missing node features by fitting \model only to the observed features. In other words, when we update the \topic membership $\phi$ or the feature model parameter $W$ by the gradient method from Equation (\ref{eq:updatephiw}) and (\ref{eq:updatew}), we only average the terms corresponding to the observed data. For example, when there is missing feature data, Equation (\ref{eq:updatephiw}) can be converted into as:
\begin{align}
\PLF{\phi_{ik}} &= \frac{\sum_{l: F_{il} \in \mathbb{O}} (F_{il} - y_{il}) w_{lk}}{\sum_{l: F_{il} \in \mathbb{O}} 1}
\end{align}
for the observed data $\mathbb{O}$.

Similarly, for link prediction we modify the model estimation method as follows. While updating the node feature model parameters $W$ based on the features of all the nodes including a new node, we estimate the network model parameters $\Theta$ only on the observed network by holding out the new node.
This way, the observed features naturally update the \topic~memberships of a new node, we can predict the missing node features or network links by using the estimated \topic~memberships and model parameters.


\section{Experiments}
\label{sec:experiments}
\newcommand{\BMAVG}{AVG}
\newcommand{\BMNBH}{NBH}
\newcommand{\BMATTR}{ATTR}
\newcommand{\BMNET}{\magfit}
\newcommand{\BMNAIVE}{CC-N}
\newcommand{\BMLOGIT}{CC-L}
\newcommand{\BMRTM}{RTM}
\newcommand{\BMMMSB}{PL-LDA}

\rev{Here} 
we perform experiments to evaluate our model. First, we run the various prediction tasks: missing node feature prediction, missing link prediction, and supervised node classification. In all tasks our model outperforms natural baselines.
Second, we qualitatively analyze the relationships between node features and network structure by a case study of a Facebook ego-network and show how the \model identifies useful and interpretable latent structures.

%

%

\xhdr{Datasets}
%
For our experiments, we used the following datasets containing networks and node features. 
%
\begin{itemize}
\denselist
\item{AddHealth (AH):} School friendship network (458 nodes, 2,130 edges) with 35 school-related node features such as GPA, courses taken, and placement~\cite{addhealth}.
%
\item{Egonet (EGO): } Facebook ego-network of a particular user (227 nodes, 6,348 edges) and 14 binary features (\eg~same high school, same age, and sports club),
\rev{manually assigned to each friend by the user.}
%
\item{Facebook100 (FB): } Facebook network of Caltech (769 nodes, 33,312 edges) and 24 university-related node features like major, gender, and dormitory~\cite{facebook100}.
%
\item{WebKB (WKB): } Hyperlinks between
computer science webpages of Cornell University in the WebKB dataset (195 nodes, 304 edges). We use occurrences of 993 words as binary features~\cite{webkb}.
\end{itemize}
\rev{We binarized discrete valued features (\eg~school year) based on whether the feature value is greater than the median value. For the non-binary categorical features (\eg~major), we used an indicator variable for each possible feature value. Some of these datasets and the source code of our algorithms are available at \textit{http://snap.stanford.edu}.}

\begin{figure}[t]
\centering
\small
\begin{tabular}{lll}
\includegraphics[width=0.13\textwidth]{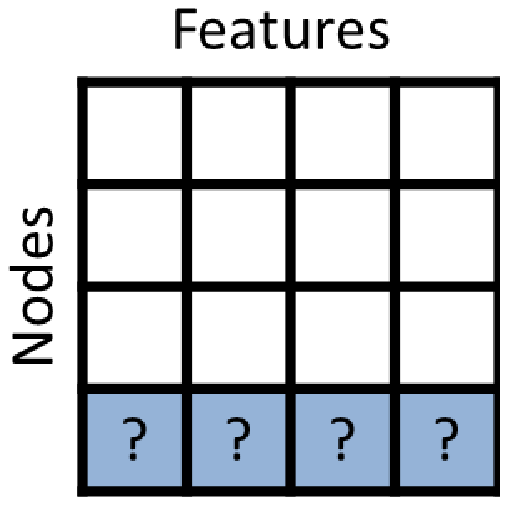} &
\includegraphics[width=0.13\textwidth]{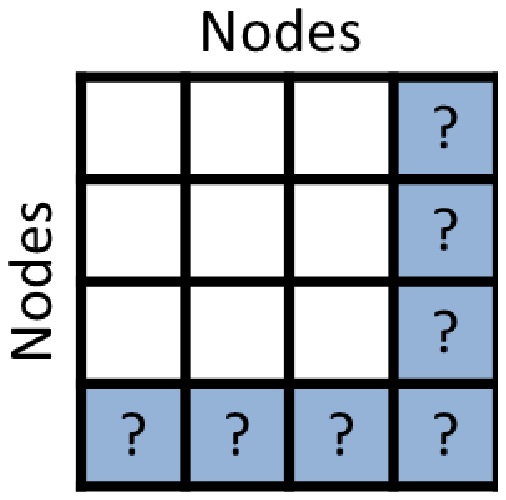} &
\includegraphics[width=0.13\textwidth]{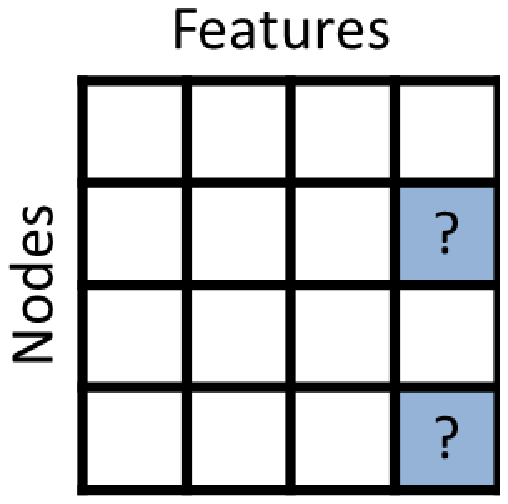} \\
(a) Missing feature & (b) Missing link & (c) Supervised node \\
~~~~prediction & ~~~~prediction & ~~~~classification\\
\end{tabular}
\caption{Three link and feature based predictive tasks.}
\vspace{-3mm}
\label{fig:predictiontask}
\end{figure}

\xhdr{Predictive tasks}
We investigate the predictive performance of the \model based on three different tasks. We visualize the three prediction tasks in Figure~\ref{fig:predictiontask}. Note that the column represents either features or nodes according to the type of the task. For each matrix, given 0/1 values in the white area, we predict the values of the entries with question marks. First, assuming that all node features of a given node are completely missing, we predict all the features based on the links of the node (Figure~\ref{fig:predictiontask}(a)). Second, when all the links of a given node are missing, we predict the missing links by using the node feature information (Figure~\ref{fig:predictiontask}(b)). Last, we assume only few features of a node are missing and we perform the supervised classification of a specific node feature given all the other node features and the network (Figure~\ref{fig:predictiontask}(c)).
%
%

\xhdr{Baseline models}
%
Now we introduce natural baseline and state of the art methods. First, for the most basic baseline model, when predicting some missing value (node feature or link) of a given node, we average the corresponding values of all the other nodes and regard it as the probability of value $1$. We refer to this algorithm as \BMAVG. Second, as we can view each of the three prediction tasks as the classification task, we use Collective Classification (CC) algorithms that exploit both node features and network dependencies~\cite{srl08ai}.
For the local classifier of CC algorithms, we use Naive-Bayes (\BMNAIVE) as well as logistic regression (\BMLOGIT). We also compare the \model to the state or the art Relational Topic Model (\BMRTM)~\cite{chang09rtm}. We give further details about these models 
and how they were applied in the \LONGER.
%
\hide{
The baseline methods which we used are as follows:
\begin{itemize}
\item{\BMAVG :} For any missing value (feature or link),
we average the corresponding values of all the other nodes and regard it as the probability that the given test node takes $1$ for the missing value.
%
\item{\BMNAIVE :} We run the Collective Classification using Naive-Bayes local classifiers.
For the aggregate function of neighbor information, we use \texttt{MODE}, which takes the majority value of the neighbors.
%
\item{\BMLOGIT :} We use the same input features as \BMNAIVE, but apply the logistic regression classifier.
Furthermore, it takes \texttt{AVERAGE} as the aggregate function of neighbor node features.
%
%
\end{itemize}
}

\xhdr{Task 1: Predicting missing node features}
First, we examine the performance for the task of predicting missing features of a node where features of other nodes and all the links are observed. We randomly select a node and remove all the feature values of that node and try to recover them.
%
%
%
We quantify the performance by using the log-likelihood of the true feature values over the estimated distributions as well as the predictive accuracy
(the probability of correctly predicting the missing features)
of each method.
%
%
\hide{
\begin{table}[t]
\centering
\small
\begin{tabular}{|l||l|l|l|l||l|l|l|l|l|}
\hline
 & \multicolumn{4}{c||}{LL} & \multicolumn{4}{c|}{ACC}\\
 \hline
 & AH & EGO & FB & WKB & AH & EGO & FB & WKB \\
 \hline
 \BMAVG & -23.0 & -5.4 & -8.7 & -179.3 & 0.53 & 0.79 & 0.77 & 0.88 \\
 \hline
 \BMNAIVE &  -17.6 &  -6.6 &  -11.6 &  -186.8 & 0.61 & 0.81 & 0.76 & 0.88 \\
 \hline
 \BMLOGIT &  -16.8 &  -5.1 &  -8.9 &  -179.2 & 0.56 & 0.78 & 0.75 & 0.89 \\
 \hline
 \BMRTM &  -xx.x &  -x.x &  -x.x &  -xxx.x & 0.xx & 0.xx & 0.xx & 0.xx \\
 \hline
 \model &  -15.6 &  -3.7 &  -7.4 &  -173.6 & 0.64 & 0.86 & 0.80 & 0.90 \\
\hline
\end{tabular}
\caption{Prediction of missing features of a node.}
\label{tbl:missattr}
\end{table}
}
\begin{table}[t]
\small
\centering
\begin{tabular}{|c||c|c|c|c|c|}
\hline
 \multicolumn{6}{|l|}{LL}\\
 \hline
 & \BMAVG & \BMNAIVE & \BMLOGIT & \BMRTM & \model \\
 \hline
 AH & -23.0 & -17.6 & -16.8 &  -63.4 & \textbf{-15.6} \\
 \hline
 EGO & -5.4 &  -6.6 &  -5.1 &  -9.9 & \textbf{-3.7} \\
 \hline
 FB & -8.7 & -11.6 & -8.9 &  -19.0 & \textbf{-7.4} \\
 \hline
 WKB &  -179.3 &  -186.8 &  -179.2 &  -336.8 & \textbf{-173.6} \\
 \hline
 \hline
 \multicolumn{6}{|l|}{ACC}\\
 \hline
 & \BMAVG & \BMNAIVE & \BMLOGIT & \BMRTM & \model \\
 \hline
 AH & 0.53 & 0.61 & 0.56 & 0.59 & \textbf{0.64} \\
 \hline
 EGO & 0.79 & 0.81  & 0.78  &  0.74 & \textbf{0.86} \\
 \hline
 FB & 0.77 & 0.76  & 0.75 & 0.77 & \textbf{0.80} \\
 \hline
 WKB &  0.88 &  0.88 &  0.89 &  0.88 & \textbf{0.90} \\
 \hline
\end{tabular}
\caption{\rev{Prediction of missing node attributes}. The \model performs the best in terms of the log-likelihood as well as the classification accuracy on the held-out data.}
\vspace{-2mm}
\label{tbl:missattr}
\end{table}

Table~\ref{tbl:missattr} shows the results of the experiments by measuring the average of log-likelihood (LL) and prediction accuracy (ACC) for each algorithm and each dataset.
We notice that \model~model exhibits the best performance in the log-likelihood for all datasets. While \BMLOGIT~in general performs the second best, our model outperforms it by up to 23\%. The performance gain over the other models in terms of accuracy seems smaller when compared to the log-likelihood. However, \model~model still predicts the missing node features with the highest accuracy on all datasets.

In particular, the \model exhibits the most improvement in node feature prediction on the ego-network dataset (30\% in LL and 7\% in ACC) over the next best method. As the node features are derived by manually labeling community memberships of each person in the ego-network dataset, a certain group of people in the network intrinsically share some node feature (community membership). In this sense, the node features and the links in the ego-network are directly related to each other and our model successfully exploits this relationship to predict missing node features.

\newcommand{\GNP}{$G_{np}$}

\xhdr{Task 2: Predicting missing links}
Second, we also consider the task of predicting the missing links of a specific node while the features of the node are given.
Similarly to the previous task, we select a node at random, but here we remove all its links while observing its features. We then aim to recover the missing links.
%
%
For evaluation, we use the log-likelihood (LL) of missing links as well as the area under the ROC curve (AUC) of missing link prediction.

\hide{
\begin{table}[t]
\small
\centering
\begin{tabular}{|l||l|l|l|l||l|l|l|l|}
\hline
 & \multicolumn{4}{c||}{LL} & \multicolumn{4}{c|}{AUC}\\
 \hline
 & AH & EGO & FB & WKB & AH & EGO & FB & WKB \\
 \hline
 \BMAVG & -40.2 & -142.7 & -320.8 & -54.2 & 0.51 & 0.61 & 0.73 & 0.70\\
 \hline
 \BMNAIVE &  -57.2 &  -134.3 & -330.7 & -185.5 & 0.69 & 0.89 & 0.70 & 0.86 \\
 \hline
 \BMLOGIT & -38.9 & -157.6 & -345.6 & -39.6 & 0.39 & 0.55 & 0.57 & 0.55 \\
 \hline
 \BMRTM &  -100.6 &  -149.9 &  -359.1 &  -25.8 & 0.56 & 0.49 & 0.46 & 0.50 \\
 \hline
 \model & -36.1 & -125.9 & -328.3 & -13.7 & 0.72 & 0.89 & 0.73 & 0.89 \\
\hline
\end{tabular}
\caption{Prediction of missing links of a node. The \model performs best in all but one case.}
\label{tbl:missnet}
\end{table}
}
\begin{table}[t]
\small
\centering
\begin{tabular}{|c||c|c|c|c|c|}
\hline
 \multicolumn{6}{|l|}{LL}\\
 \hline
 & \BMAVG & \BMNAIVE & \BMLOGIT & \BMRTM & \model \\
 \hline
 AH & -40.2 & -57.2 & -38.9 & -100.6 & \textbf{-36.1} \\
 \hline
 EGO & -142.7 &  -134.3 &  -157.6 &  -149.9 & \textbf{-125.9} \\
 \hline
 FB & \textbf{-320.8} & -330.7 & -345.6 & -359.1  & -328.3 \\
 \hline
 WKB &  -54.2 &  -185.5 &  -39.6 &  -25.8 & \textbf{-13.7} \\
 \hline
 \hline
 \multicolumn{6}{|l|}{AUC}\\
 \hline
 & \BMAVG & \BMNAIVE & \BMLOGIT & \BMRTM & \model \\
 \hline
 AH & 0.51 & 0.69 & 0.39 & 0.56 & \textbf{0.72}\\
 \hline
 EGO & 0.61 & \textbf{0.89}  & 0.55  &  0.49 & \textbf{0.89} \\
 \hline
 FB & \textbf{0.73} & 0.70  & 0.57 & 0.46 & \textbf{0.73} \\
 \hline
 WKB &  0.70 &  0.86 &  0.55 & 0.50  & \textbf{0.89} \\
 \hline
\end{tabular}
\caption{Prediction of missing links of a node. The \model performs best in all but one case.}
\vspace{-2mm}
\label{tbl:missnet}
\end{table}

We give the experimental results for each dataset in Table~\ref{tbl:missnet}. Again, the \model outperforms the baseline models in the log-likelihood except for the Facebook100 data.
Interestingly, while \BMRTM~was relatively competitive when predicting missing features, it tends to fail predicting missing links, which \rev{implies} that the flexibility of link-affinity matrices is needed for accurate modeling of the links.

\rev{We observe that}
Collective Classification methods 
\rev{look competetive}
in some performance metrics and datasets. For example, \BMNAIVE~gives good results in terms of classification accuracy,  and \BMLOGIT~performs well in terms of the log-likelihood.
As \BMNAIVE~is a discriminative model, it does not perform well in missing link probability estimation. However, the \model is a generative model that produces a joint probability of node features and network links, so it is also very good at estimating missing links. Hence, in overall, the \model nicely exploits the relationship between the network structure and node features to predict missing links.

\xhdr{Task 3: Supervised node classification}
Finally, we examine the performance on the supervised classification task.
In many cases, we aim to classify entities (nodes) based on their feature values under the supervised setting. Here the relationships (links) between the entities are also provided. For this experiment, we hold out one feature of nodes as the output class, 
\rev{regarding} all other features of nodes and the network as input data. 
We divide the nodes into a 70\% training and 30\% test set.
Similarly, we measure the average of the log-likelihood (LL) 
as well as the average classification accuracy (ACC) on the test set.
%

\hide{
\begin{table}[t]
\small
\centering
\begin{tabular}{|l||l|l|l|l||l|l|l|l|}
\hline
 & \multicolumn{4}{c||}{LL} & \multicolumn{4}{c|}{ACC}\\
 \hline
 & AH & EGO & FB & WKB & AH & EGO & FB & WKB \\
 \hline
 \BMAVG & -84.5 & -24.8 & -97.6 & -17.5 & 0.52 & 0.76 & 0.69 & 0.82 \\
 \hline
 \BMNAIVE &  -486.6 &  -54.0 &  -254.6 & -290.4 & 0.58  & 0.76 & 0.71 & 0.81 \\
 \hline
 \BMLOGIT & -60.5 & -22.2 & -79.2 & -15.4 & 0.63 & 0.77 & 0.77 & 0.84 \\
 \hline
 \BMRTM &  &  &  &  & 0.xx & 0.xx & 0.xx & 0.xx \\
 \hline
 \model & -55.3 & -21.2 & -63.4 & -15.0 & 0.63 & 0.79 & 0.81 & 0.85 \\
\hline
\end{tabular}
\caption{Supervised node classification.}
\label{tbl:missclass}
\end{table}
}
\begin{table}[t]
\small
\centering
\begin{tabular}{|c||c|c|c|c|c|}
\hline
 \multicolumn{6}{|l|}{LL}\\
 \hline
 & \BMAVG & \BMNAIVE & \BMLOGIT & \BMRTM & \model \\
 \hline
 AH & -84.5 & -486.6 & -60.5 &  -236.0 & \textbf{-55.3} \\
 \hline
 EGO & -24.8 &  -54.0 &  -22.2 &  -41.7 & \textbf{-21.2} \\
 \hline
 FB & -97.6 & -254.6 & -79.2 &  -181.7 & \textbf{-63.4} \\
 \hline
 WKB &  -17.5 &  -254.6 &  -15.4 &  -193.6 & \textbf{-15.0} \\
 \hline
 \hline
 \multicolumn{6}{|l|}{ACC}\\
 \hline
 & \BMAVG & \BMNAIVE & \BMLOGIT & \BMRTM & \model \\
 \hline
 AH & 0.52 & 0.58 & \textbf{0.63} & 0.51 & \textbf{0.63} \\
 \hline
 EGO & 0.76 & 0.76  & 0.77  &  0.75 & \textbf{0.79} \\
 \hline
 FB & 0.69 & 0.71  & \textbf{0.77} & 0.72 & \textbf{0.77} \\
 \hline
 WKB &  0.82 &  0.81 &  0.84 &  0.84 & \textbf{0.85} \\
 \hline
\end{tabular}
\caption{Supervised node classification. The \model gives the best performance on both metrics and all four datasets.}
\label{tbl:missclass}
\end{table}


We illustrate the performance of various models in Table~\ref{tbl:missclass}. The \model~model performs better than the other models in both the log-likelihood and the classification accuracy. It improves the performance by up to 20\% in the log-likelihood and 5\% in the classification accuracy. We also notice that exploiting the relationship between node features and global network structure can improve the performance on supervised node classification compared to the models focusing on the local network dependencies (\eg, Collective Classification methods).

\hide{
To take a look at the results in detail,
we can also find that the network is highly related to the classification task in the ego-network,
since \BMNBH~and \model~shows better performance than the others.
Similarly to the first experiment, using the other features and the whole network structure in the \model~model brings performance improvement over~\BMNBH.
Moreover, we can see that the logistic regression shows poor performance in the ego-network compared to the other datasets.
We interpret this as follows.
Since the network links are hard to be represented by linear combination of node features~\cite{mh11magfit},
the logistic regression shows poor performance as the network information becomes important.

In conclusion, from the above results,
we can see that adding network information can help with improving the classification performance.
}

\newcommand{\GFormat}[1]{\textsc{#1}}
\newcommand{\GKorComp}{\GFormat{KComp}}
\newcommand{\GHS}{\GFormat{HS}}
\newcommand{\GBasket}{\GFormat{BasketBall}}
\newcommand{\GSquash}{\GFormat{Squash}}
\newcommand{\GBS}{\GFormat{University}}
\newcommand{\GFam}{\GFormat{Family}}
\newcommand{\GCS}{\GFormat{CS}}
\newcommand{\GAge}{\GFormat{Age}}
\newcommand{\GEbay}{\GFormat{Intern}}
\newcommand{\GKorClub}{\GFormat{KProg}}
\newcommand{\GStan}{\GFormat{ST}}
\newcommand{\GTravel}{\GFormat{Travel}}
\newcommand{\GKorStan}{\GFormat{KorST}}
\newcommand{\GMath}{\GFormat{Camp}}

\xhdr{Case study: Analysis of a Facebook ego-network}
Now we qualitatively analyze the Facebook ego-network example 
\rev{to provide}
insights into the relationship between node features and network structure. We examine the estimated model parameters $W$ (for features) and $\Theta$ (for network structure). By investigating model parameters ($W$ and $\Theta$), we can find not only what features are important for each \topic but also how each \topic affects the link structure.

We begin by introducing the user which we used to create a network between his Facebook friends. We asked our user to label each of his friends with a number of labels. He chose to use 14 different labels. They correspond to his high school (\GHS), undergraduate university (\GBS), math olympiad camp (\GMath), computer programming club (\GKorClub) and  work place (\GKorComp) friends. The user also assigned labels to identify friends from his graduate program (\GCS) and university (\GStan), basketball (\GBasket) and squash (\GSquash) clubs, as well as travel mates (\GTravel), summer internship buddies (\GEbay), family (\GFam) and age group (\GAge).
\hide{
Table~\ref{tbl:egodesc} summarizes each community and its brief description of the ego-network.
\begin{table*}[t]
\centering
\begin{tabular}{|l|l||l|l|}
\hline
Group & Description & Group & Description\\
\hline
\hline
\GHS & High school & \GBS & University for BS\\
\hline
\GStan & Stanford & \GCS & Stanford Computer Science\\
\hline
\GKorStan & Korean at Stanford & \GKorClub &  Club at Korean university\\
\hline
\GBasket & First Club at Stanford & \GSquash & Second Club at Stanford\\
\hline
\GKorComp & Korean company & \GEbay & Company for the summer internship\\
\hline
\GMath & Math camp & \GTravel & Friends who go travels together\\
\hline
\GAge & Friends of the samge age & \GFam & Family\\
\hline
\end{tabular}
\caption{Description of communities in the Facebook ego-network}
\label{tbl:egodesc}
\end{table*}
}
%
%

\rev{We fit the \model to} 
the ego-network and each friend's memberships to the above communities. We obtained the model parameters $W$ and $\Theta$.
For the validation procedure, we set the number of latent groups to $5$ since the previous prediction tasks worked well when $K = 5$.
In Table~\ref{tbl:egow}, for each of $5$ latent groups, we represent the top 3 features with the largest absolute value of model parameter $|w_{lk}|$ and the corresponding link-affinity matrices $\Theta_{k}$.
\hide{
Note that, by the model description in Equation~(\ref{eq:modelattr})
the \topic~membership $\phi_{ik}$ and the corresponding feature $F_{il}$ have high positive corrleation as the model parameter $w_{lk}$ is large.
In other words, when the node $i$ belongs to the $k$-th \topic~with high probability ($\phi_{ik}$ is large),
the node feature $F_{il}$ tends to take value $1$ in this case.
Conversely, if $w_{lk}$ is small ($w_{lk}$ can be less than 0), then the node feature $F_{il}$ is likely to be $0$ when the node $i$ belongs to the $k$-th \topic~with high probability.
Similarly, the negligible value of $w_{lk}$  (\ie , $|w_{lk}| \approx 0$) indicates that the $l$-th node feature is almost independent of the $k$-th \topic .

On the other hand, in our notation of the link-affinity matrix, each entry indicates the relative link-affinity between a pair of nodes depending on their memberships to a given \topic .
For example, the last entry (1, 1) in each matrix indicates the link-affinity between a pair of nodes when the both nodes belong to the specified \topic .
}

We begin by investigating the first \topic. The top three labels the most correlated to the first \topic are \GStan ,  \GAge , and \GEbay.
However, notice that \GEbay~is negatively correlated.
This means that \topic 1 contains students from the same graduate school and age, but not people with whom our user worked together at the summer internship (even though they may be of the same school/age). We also note that $\Theta_{1}$ exhibits homophily structure.
From this we learn that summer interns, who met our Facebook user neither because of shared graduate school nor because of the age, form a group within which people are densely connected. On the other hand, people of the same age at the same university also exhibit the homophily, but are less densely connected with each other. Such variation in link density 
that depends 
on the group memberships agrees with  our intuition. Those who worked at the same company actively interact with each other so almost everyone is 
\rev{linked}
in Facebook. However, 
\rev{as the group of people of the same university or age}
is large and each pair of people in that group does not necessarily know each other, the link affinity in this group is naturally smaller than in the intern's group.

\hide{
In addition, we are able to notice similar phenomena in the groups 3 and 4.
The \topic~3 has negative correlation with \GHS~and \GBS , while it shows positive correlation with \GBasket. Even in the groups with negative correlation, \GHS is more strongly correlated with \topic~3 than \GBS .
Such correlations imply as follows.
When the membership of \topic~3 separates people into two groups,
the non-members of \topic~3 are likely to be in \GHS~and \GBS
whereas the members of it tends to be in \GBasket .
Furthermore, according to the link-affinity matrix of \topic~3,
the link-affinity is the highest between members of \topic~3,
the second highest between non-members,
and that of the other case (between a member and a non-member) follows.
Hence, in overall, we demonstrate that people in \GBasket~form a fairly strong community and people in \GHS~and \GBS~also form a community.
Similarly, we figure out that people in \GBS~forms a community structure.
}

Similarly, groups 2 and 3 form the two sports groups (\GBasket , \GSquash). People are connected densely within each of the groups, but less connected to the outside of the groups. This is natural because the sports clubs make members actively interact with each other but do not necessarily make members interact with those not in the clubs.
%
\rev{Furthermore, we}
notice that those who graduated from not only the same high school \rev{(\GHS)} but also the same undergraduate school \rev{(\GBS)} form another community but the membership to high school is more important than to the undergraduate university (8.7 vs. 2.3).

However, for groups 4 and 5, we note that the corresponding link-affinity matrices are nearly flat (\ie~values are nearly uniform).
This implies that groups 4 and 5 are related to general node features. In this sense, we hypothesize that features like CS, family, math camp, and the company, have relatively little effect on the network structure.

\begin{table*}[t]
\small
\centering
\begin{tabular}{|l||l|l|l||l|}
\hline
Group & Top 1 & Top 2 & Top 3 & Link-affinity matrix\\
\hline
\hline
1 & \GStan~(9.0) & \GAge~(4.5) &   \GEbay~(-3.7) & [0.67~0.08;~0.08~0.17]\\
\hline
2 & \GHS~(-8.7) & \GBS~(-2.3) &  \GBasket~(2.2)  & [0.26~0.18;~0.18~0.38]\\
\hline
3 & \GBS~(-7.1) & \GKorStan~(-2.6) & \GSquash~(2.2)  & [0.22~0.23;~0.23~0.32]\\
\hline
4 & \GCS~(7.3) &  \GFam~(7.0) &   \GMath~(6.9) & [0.25~0.24;~0.24~0.27]\\
\hline
5 & \GKorComp~(5.2) &  \GKorStan~(4.4) &  \GEbay~(-3.8) & [0.29~0.22;~0.22~0.27]\\
\hline
\end{tabular}
\caption{\rev{Logistic model parameter values of top 3 features and the link-affinity matrix associated with each \topic~in the ego-network.}}
\label{tbl:egow}
\vspace{-4mm}
\end{table*}

\section{Related Work and Discussion}
The \model builds on previous research in machine learning and network analysis. Many models have been developed to explain network link structure \cite{airoldi07blockmodel,hoff02latent,kemp06irm,jure10kronecker} and extensions that incorporate node features have also been proposed~\cite{getoor01,mh11magfit,taskar03linkpred}. However, these models do not consider latent groups 
and thus 
cannot provide the benefits of dimensionality reduction \rev{or} produce interpretable clusters useful for understanding network community structure.


The \model provides meaningful clustering of nodes and \rev{their} features \rev{in the network}. 
\rev{The network models of similar flavor}
have been proposed in the past~\cite{airoldi07blockmodel,hoff02latent,kemp06irm}, and some even incorporate node features~\rev{\cite{chang09rtm,nallapati08lda,miller09nfrm}}. However, such models have been mainly developed for document networks 
\rev{where they}
assume the multinomial 
\rev{topic distributions for each word in the document.}
We extend this by learning a logistic model for occurrence of each feature based on node group memberships. 
\rev{To highlight the difference between the previous models and ours, since}
topic memberships in the above models are modeled by multinomial distributions,
a node \rev{has} a mass of 1 to split among various topics. In contrast, 
in the \model, 
a node can belong to multiple topics at once \rev{without any constraint}.

While previous work tends to explore only the network or only the features, the \model jointly models both
\rev{so that it can}
make predictions on one given the other.  The \model models the interaction between links and group memberships 
\rev{via}
link-affinity matrices which 
provide great flexibility and interpretability of obtained groups and interactions.

The \model is a new probabilistic model of links and nodes in networks. It can be used for link prediction, node feature prediction and supervised node classification. We have demonstrated qualitatively and quantitatively that the \model proves useful for analyzing network data. The \model significantly improves on previous models, integrating both node-specific information and link structure to give better predictions.


\section*{Acknowledgments}
\rev{
Myunghwan Kim was supported by the Kwanjeong Educational Foundation fellowship.
This research has been supported in part by NSF
CNS-1010921,                
IIS-1016909,                
IIS-1149837,        
IIS-1159679,                
Albert Yu \& Mary Bechmann Foundation, Boeing, Allyes, Samsung, Yahoo,
Alfred P. Sloan Fellowship and the Microsoft Faculty Fellowship.
}


\bibliography{icml-maglda}
\bibliographystyle{icml2012}

\clearpage
\appendix
\section{Mathematical Details}

\subsection{Update of Group Membership $\phi$}
In Equation~(\ref{eq:updatephiw}), we proposed the gradient ascent method which updates each \topic~membership $\phi_{ik}$ to maximize the lower bound of log-likelihood $\LZ$.
To complete its computation,
we further take a look at $\frac{\partial \EZPH \log p_{ij}}{\partial \phi_{ik}}$ and $\frac{\partial \EZPH \log (1-p_{ij})}{\partial \phi_{ik}}$ in detail.
Then, we can also compute $\frac{\partial \EZPH \log p_{ji}}{\partial \phi_{ik}}$ and $\frac{\partial \EZPH \log (1-p_{ji})}{\partial \phi_{ik}}$ in the same way.

First, we calculate the derivative of expected log-likelihood for edges, $\EZPH \log p_{ij}$.
When all the \topic~memberships except for $\phi_{ik}$ are fixed, 
we can derive $\frac{\partial \EZPH \log p_{ij}}{\partial \phi_{ik}}$ 
from definition of $p_{ij}$ in Equation~(\ref{eq:modelnet})
as follows:
\begin{align}
\label{eq:preupdatephiattr1}
\frac{\partial \EZPH \log p_{ij}}{\partial \phi_{ik}} 
& = \frac{\partial}{\partial \phi_{ik}} \EZPH \left[ \sum_{k'} \log \Theta_{k'} [z_{ik'}, z_{jk'} ] \right] \nonumber \\
& = \sum_{k'} \left[ \frac{\partial}{\partial \phi_{ik}} \EZPH \log \Theta_{k'} [z_{ik'}, z_{jk'} ] \right]
\end{align}
Here we use the following property.
Since $z_{ik}$ is an independent Bernoulli random variable with probability $\phi_{ik}$, 
for any function $f : \{0, 1\}^{2} \rightarrow \mathbb{R}$,
\begin{align}
\label{eq:expectation}
\EZPH f(z_{ik}, z_{jk}) = \phi_{ik}\phi_{jk}f(1,1) + \phi_{ik}(1-\phi_{jk})f(1, 0) \nonumber \\
\quad + (1-\phi_{ik})\phi_{jk}f(0, 1) + (1-\phi_{ik})(1-\phi_{jk})f(0, 0) \, .
\end{align}
Hence, by applying Equation~(\ref{eq:expectation}) to (\ref{eq:preupdatephiattr1}), we obtain
\begin{align}
\label{eq:updatephiattr1}
& \frac{\partial \EZPH \log p_{ij}}{\partial \phi_{ik}} 
= \frac{\partial}{\partial \phi_{ik}} \EZPH \log \Theta_{k} [z_{ik}, z_{jk} ]  \nonumber \\
& = \phi_{jk} \log \Theta_{k}[1, 1] + (1-\phi_{jk}) \log \Theta_{k}[1, 0] \quad \quad \quad \nonumber \\
& \quad \quad - \phi_{jk} \log \Theta_{k}[0, 1] - (1-\phi_{jk}) \log \Theta_{k} [0, 0] \, .
\end{align}

Next, we compute the derivative of expected log-likelihood for unlinked node pairs, \ie~$\EZPH \log (1 - p_{ij})$.
Here we approximate the computation using the Taylor's expansion, $\log (1 - x) \approx - x - 0.5 x^{2}$ for small $x$:
\begin{align}
\frac{\partial \EZPH \log (1 - p_{ij})}{\partial \phi_{ik}}
& \approx - \frac{\partial \EZPH  p_{ij}}{\partial \phi_{ik}}
- 0.5 \frac{\partial \EZPH  p_{ij}^{2} }{\partial \phi_{ik}} \, . \nonumber
\end{align}
To compute $\frac{\partial \EZPH p_{ij}}{\partial \phi_{ik}}$,
\begin{align}
& \frac{\partial \EZPH p_{ij}}{\partial \phi_{ik}}  \nonumber \\
& = \frac{\partial}{\partial \phi_{ik}} \EZPH \prod_{k'} \Theta_{k'}[z_{ik'}, z_{jk'}] \nonumber \\
& = \frac{\partial}{\partial \phi_{ik}} \EZPH \Theta_{k}[z_{ik}, z_{jk}] \prod_{k' \neq k} \Theta_{k'}[z_{ik'}, z_{jk'}] \nonumber \\
& = \prod_{k' \neq k} \EZPH \Theta_{k'}[z_{ik'}, z_{jk'}] \frac{\partial}{\partial \phi_{ik}} \EZPH \Theta_{k}[z_{ik}, z_{jk}] \nonumber \, .
\end{align}
By Equation~(\ref{eq:expectation}), each $\EZPH \Theta_{k}[z_{ik}, z_{jk}]$ and its derivative can be obtained.
Similarly, we can calculate $\frac{\partial \EZPH p^{2}_{ij}}{\partial \phi_{ik}}$, so we complete the computation of $\frac{\partial \EZPH \log (1-p_{ij})}{\partial \phi_{ik}}$.
%
%
%
%
%

As we attain $\frac{\partial \EZPH \log p_{ij}}{\partial \phi_{ij}}$ and $\frac{\partial \EZPH \log (1-p_{ij})}{\partial \phi_{ij}}$,
we eventually calculate $\PLA{\phi_{ik}}$.
Hence, by adding up $\PLPH{\phi_{ik}}$, $\PLF{\phi_{ik}}$, and $\PLA{\phi_{ik}}$, we complete computing the derivative of the lower bound of log-likelihood $\PLZ{\phi_{ik}}$:
\begin{align*}
\PLZ{\phi_{ik}} = \PLA{\phi_{ik}} + \PLF{\phi_{ik}} + \PLA{\phi_{ik}} \, .
\end{align*}

\hide{
Finally, we update the \topic~membership $\phi_{ik}$ using the gradient ascent method:
\begin{align}
\phi_{ik}^{new} = \phi_{ik}^{old} + \gamma_{\phi} \PLZ{\phi_{ik}} \nonumber
\end{align}
for a given learning rate $\gamma_{\phi}$.
By updating each $\phi_{ik}$ in turn with fixing the others,
we can find the optimal \topic~membership $\phi$ given the model parameters $W$ and $\Theta$.
}

\hide{
\subsection{Update of Logistic Model Parameters $W$}
Now we turn our attention to the update of parameters for the node attribute model, $W$, while every \topic~membership $\phi_{ik}$ for each node is fixed.
Note that given the \topic~membership $\phi$ the node attribute model (logistic) and the network model (MAG) are independent of each other.
Therefore, 
finding the parameter $W$ is identical to fitting the logistic model to those given input ($\phi$) and output ($F$) data,
because $\LPH$ and $\LA$ are irrelevant to $W$.
Furthermore, as we penalize the objective function in Equation~(\ref{eq:problem2}) on the L1 value of the model parameter $W$,
finding the optimal $W$ given $\phi$ can be viewed as a L1-regularized logistic regression.

Here we use the L1-regularized gradient method to update the parameter $W$.
To proceed, we need to compute $\PLF{w_{lk}}$ and $\PLF{\mu_{l}}$ for each $l = 1, \cdots, L$ and $k = 1, \cdots, K$.
\begin{align}
\label{eq:updatew}
\PLF{w_{lk}} & = \sum_{i} (F_{il} - y_{il}) \phi_{ik} \nonumber \\
\PLF{\mu_{l}} & = \sum_{i} (F_{il} - y_{il})
\end{align}
First, we update $\mu_{l}$ without consideration of L1 penalty as follows:
\begin{align}
\PLF{\mu_{l}}^{new} = \mu_{l}^{old} + \gamma_{F} \PLF{\mu_{l}} \nonumber
\end{align}
for the constant learning rate $\gamma_{F}$.

Next, we consider the L1-regularization for each $w_{lk}$.
When we define $\mathcal{A} = \{w_{lk} : w_{lk} \neq 0 \}$ as the active set,
our purpose is to make $\mathcal{A}$ sparse.
The main idea for the sparse active set is not to allow the small absolute value of $w_{lk}$.
To be concrete, when $w_{lk} \in \mathcal{A}$ (\ie~$w_{lk} = 0$) and $|\frac{\partial \LF}{\partial w_{lk}}| < \lambda$,
we do note update $w_{lk}$ at all.
Moreover, whenever $w_{lk}$ changes its sign while being updated,
we make the $w_{lk}$ inactive, \ie~remove the $w_{lk}$ from the active set $\mathcal{A}$.

To describe formally, we update $w_{lk}$ by
\[
w_{lk}^{new} = w_{lk}^{old} + \gamma_{F} \PLF{w_{lk}} - \lambda \mathrm{Sign}(w_{lk})
\]
only if $w_{lk} \neq 0$ or $\gamma_{F} | \PLF{w_{lk}} | > \lambda$.
After the update, if $\mathrm{Sign}(w_{lk}^{new}) \neq \mathrm{Sign}(w_{lk}^{old})$,
then we reassign $w_{lk}^{new} = 0$, \ie~make the $w_{lk}$ inactive.
By this procedure, we can update the node attribute model parameter $W$
to maximize the lower bound of log-likelihood $\LZ$
as well as
to maintain the small number of relevant \topic s for each node attribute.
}

\subsection{Update of MAG Model Parameters $\Theta$}
Next we focus on the update of parameters of the network model, $\Theta$, where the \topic~membership $\phi$ is fixed.
Since the network model is independent of the node attribute model given the \topic~membership $\phi$,
we do not need to consider $\LPH$, $\LF$, or $|W|_{1}$.
We thus update $\Theta$ to maximize only $\LA$ given $\phi$ using the gradient method.

As we previously did in computing $\PLA{\phi_{ik}}$ by separating edge and non-edge terms,
we compute each $\PLA{\Theta_{k}[x_{1}, x_{2}]}$ for $k = 1, \cdots, K$ and $x_{1}, x_{2} \in \{0, 1\}$.
To describe mathematically,
\begin{align}
\label{eq:partialtheta}
\PLA{\Theta_{k}[x_{1}, x_{2}]} 
& = \sum_{A_{ij} = 1} \frac{\partial \EZPH \log p_{ij}}{\partial \Theta_{k}[x_{1}, x_{2}]} \nonumber \\
& \quad + \sum_{A_{ij} = 0} \frac{\partial \EZPH \log (1-p_{ij})}{\partial \Theta_{k}[x_{1}, x_{2}]} \, .
\end{align}

Now we compute each term in the above calculation by the definition of $p_{ij}$.
First, we compute the former term by using Equation~(\ref{eq:expectation})
For instance,
\[
\PLA{\Theta_{k}[0, 1]} = (1-\phi_{ik})\phi_{jk} \frac{\partial \log \Theta_{k}[0, 1]}{\partial \Theta_{k}[0, 1]} = \frac{(1-\phi_{ik})\phi_{jk}}{\Theta_{k}[0, 1]} \, .
\]
Hence, we can properly compute Equation~(\ref{eq:partialtheta}) depending on the values of $x_{1}$ and $x_{2}$.

Second, we use the same Taylor's expansion technique for the latter term in Equation~(\ref{eq:partialtheta}) as follows:
\begin{align}
\frac{\partial \EZPH \log(1-p_{ij})}{\partial \Theta_{k} [x_{1}, x_{2}]}
& \approx \frac{\partial}{\partial \Theta_{k}[x_{1}, x_{2}]}\EZPH \left( - p_{ij} - 0.5 p_{ij}^{2} \right) \, . \nonumber 
\end{align}
%
Similarly to $\frac{\EZPH p_{ij}}{\partial \phi_{ik}}$,
$\frac{\EZPH p_{ij}}{\partial \Theta_{k}[x_{1}, x_{2}]}$ is computed by
\[
\prod_{k' \neq k} \EZPH \Theta_{k'}[z_{ik'}, z_{jk'}] \frac{\partial}{\partial \Theta_{k}[x_{1}, x_{2}]} \EZPH \Theta_{k}[z_{ik}, z_{jk}]
\]
where each term is obtained by Equation~(\ref{eq:expectation}).
Similarly, we compute $\frac{\EZPH p^{2}_{ij}}{\partial \Theta_{k}[x_{1}, x_{2}]}$ so that we can obtain $\PLA{\Theta_{k}[x_{1} , x_{2} ]}$.

\hide{
In consequence, we can compute the gradient for each $\Theta_{k}$ and update it to maximize the lower bound $\LA$ by the gradient method:
\begin{align}
\Theta_{k}^{new} = \Theta_{k}^{old} + \gamma_{A} \nabla_{\Theta_{k}} \LA \nonumber
\end{align}
for a constant learning rate $\gamma_{A}$.
}

\section{Implementation Details}
\subsection{Initialization}
Since the objective function in Equation~(\ref{eq:problem2}) is non-convex, the final solution might be dependent on the initial values of $\phi$, $W$, and $\Theta$.
For reasonable initialization, as the node attributes $F$ are given, we run the Singular Vector Decomposition (SVD) by regarding $F$ as an $N \times L$ matrix
and obtain the singular vectors corresponding to the top $K$ singular values.
By taking the top $K$ components, we can approximate the node attributes $F$ over $K$ latent dimensions.
We thus assign the $l$-th entry of the $k$-th right singular vectors multiplied by the $k$-th singular value into $w_{lk}$ for $l = 1, \cdots, L$ and $k = 1, \cdots, K$.
We also initialize each \topic~membership $\phi_{ik}$ based on the $i$-th entry of the $k$-th left singular vectors.
This approximation can in particular provide good enough initial values
when the top $K$ singular values dominate the others.
In order to obtain the sparse model parameter $W$,
we reassign $0$ to $w_{lk}$ of small absolute value such that $|w_{lk}| < \lambda$.

Finally, to initialize the link-affinity matrices $\Theta$, we introduce the following way.
When initializing the $k$-th link-affinity matrix $\Theta_{k}$,
we assume that the \topic~other than \topic~$k$ has nothing to do with network structure, \ie~every entry in the other link-affinity matrices has the equal value.
Then, we compute the ratio between entries $\Theta_{k}[x_{1}, x_{2}]$ for $x_{1}, x_{2} \in \{0, 1\}$ as follows:
\begin{align*}
\Theta_{k}[x_{1}, x_{2}] \propto \sum_{i, j: A_{ij} = 1} \EZPH P[z_{ik} = x_{1}, z_{ik} = x_{2}]
\end{align*}
As the \topic~membership $\phi$ is initialized above and $z_{ik}$ and $z_{jk}$ are independent of each other,
we are able to compute the ratio between entries of $\Theta_{k}$.
After computing the ratio between entries for each link-affinity matrix, we adjust the scale of the link-affinity matrices 
so that the expected number of edges in the MAG model is equal to the number of edges in the given network, \ie~$\sum_{i, j} p_{ij} = \sum_{i, j} A_{ij}$.

\subsection{Selection of the Number of Groups $K$}
Another issue in fitting the \model to the given network and node feature data is to determine the number of groups, $K$.
We can find the insight about the value of $K$ from the MAG model.
It has been already proved that, in order for the MAG model to reasonably represent the real-world network, the value of $K$ should be in the order of $\log N$ where $N$ represents the number of nodes in the network~\cite{mh12mag}.
Since in the \model the network links are modeled similarly to the MAG model,
the same argument on the number of groups $K$ still holds.

However, the above argument cannot determine the specific value of $K$.
To select one value of $K$, we use the cross-validation method as follows.
For instance, suppose that we aim to predict all the features of a node where its links to the other nodes are fully observed (Task 1 in Section~\ref{sec:experiments}).
While holding out the test node, we can set up the same prediction task in a way that we select one at random from the other nodes (training nodes) and regard it as the validation test node.
We then perform the missing node feature prediction on this validation node and obtain the log-likelihood result.
By running this procedure with varying the validation test node, we can attain the average log-likelihood on the missing node features given the specific value of $K$ (\ie~N-fold cross-validation).
Finally, we compare the average log-likelihood values according the value of $K$ and pick up the best one to maximize the log-likelihood.
This method can be done by the other prediction tasks, missing link prediction and supervised node classification.

\subsection{Baseline Models}
Here we briefly describe how we implemented each baseline method depending on the type of prediction task.

\xhdr{\BMAVG}
In this baseline method, we regard each $l$-th node feature and a link to the $i$-th node as an independent random variable, respectively.
In other words, we assume that missing node features or links do not depend on each other.
Hence, we predict the $l$-th missing node feature by finding the probability that the $l$-th node feature of all the other nodes have value $1$.
We then regard the found probability as that of the missing $l$-th node feature taking value $1$.

Similarly, when we predict missing links (in particular, the link to the $i$-th node) of a given node,
we average the probability that all the other nodes are linked to the $i$-th node and take it as the probability of link from the given node to the $i$-th node
(\ie~preferential attachment).

\xhdr{\BMNAIVE}
For this method, we basically use the Naive-Bayes method using node features of each node as well as those of neighboring nodes.
To represent each node feature of neighboring nodes by a single value,
we select the majority value (either 0 or 1) from the neighbors' feature values.

However, we cannot use the node features when predicting all the node features of a given node.
Furthermore, the node features of neighboring nodes are unattainable when we predict missing links.
Therefore, depending on the type of prediction task, we exploit only achievable information among node features and those of neighboring nodes.

\xhdr{\BMLOGIT}
We employ the similar approach to the \BMNAIVE . 
However, here we use the logistic regression rather than the Naive-Bayes and average the feature values of neighboring nodes rather than pick up the majority value.

\xhdr{\BMRTM}
We use the lda-R package to run \BMRTM~(\textit{http://cran.r-project.org/web/packages/lda/index.html}).

\end{document}